\documentclass[twocolumn,showpacs,preprintnumbers,amsmath,amssymb,prl,floatfix]{revtex4}
\usepackage{graphicx}
\usepackage{dcolumn}
\usepackage{bm}
\usepackage{longtable}
\usepackage[usenames]{color}
\begin{document}
\newcommand\T{\rule{0pt}{2.6ex}}
\newcommand{\alert}[1]{\textcolor{red}{[\textrm{#1}]}}
\title{Improving the efficiency of ultracold dipolar molecule formation by first loading onto an optical lattice}
\author{J.~K.~Freericks}
\affiliation{Department of Physics, Georgetown University, Washington, D.C. 20057}
\author{M.~M.~Ma\'ska}
\affiliation{Department of Theoretical Physics, Institute of Physics, University of Silesia, 40-007 Katowice, Poland}
\author{Anzi Hu}
\affiliation{Joint Quantum Institute, NIST and University of Maryland,  100 Bureau Drive, Stop 8423, Gaithersburg Maryland 20899-8423, USA}
\author{R.~Lema\'nski}
\affiliation{Institute of Low Temperature and Structure Research, Polish Academy of Science, 50-422 Wroc{\l}aw, Poland}
\author{Thomas M. Hanna}
\affiliation{Joint Quantum Institute, NIST and University of Maryland,  100 Bureau Drive, Stop 8423, Gaithersburg Maryland 20899-8423, USA}		
\author{C. J. Williams}
\affiliation{Joint Quantum Institute, NIST and University of Maryland,  100 Bureau Drive, Stop 8423, Gaithersburg Maryland 20899-8423, USA}
\author{P. S. Julienne}
\affiliation{Joint Quantum Institute, NIST and University of Maryland,  100 Bureau Drive, Stop 8423, Gaithersburg Maryland 20899-8423, USA}
\date{\today}

\begin{abstract}Ultracold ground state dipolar $^{40}$K$^{87}$Rb molecules have recently been produced in a loose harmonic trap by employing a magnetic field sweep across a Feshbach resonance followed by stimulated Raman adiabatic passage [K.-K. Ni \textit{et al.}, Science {\bf 322}, 231 (2008)]. The overall experimental efficiency for molecule formation was around 20\%.  We show that the efficiency can be increased to nearly 100\% if one first loads the atomic gases into an optical lattice of the appropriate depth and tunes the interspecies attraction to have exactly one atom of each species at an occupied lattice site.  Our proposed scheme  provides a large enhancement to the dipolar molecule density even at relatively high temperatures, and avoids three-body recombination loss by preventing lattice sites from being triply occupied.
\end{abstract}


\maketitle

In recent years there has been intense activity in the field of ultracold molecules~\cite{carr_review}. With their long range, directional interactions and periodic structure, dipolar molecules in optical lattices have attracted particular attention. Applications of this system include the implementation of novel condensed matter phases~\cite{goral02}, quantum computation~\cite{demille02} and simulation~\cite{micheli06} schemes, and assisting the creation of a dipolar superfluid~\cite{damski03}. 
Motivated by these possibilities, a number of recent experiments have studied molecules of various isotopic combinations of K and Rb~\cite{ospelkaus06, zirbel08prl, zirbel08pra, weber08}, leading to the creation of $^{40}$K$^{87}$Rb molecules in their absolute ground state~\cite{jila_expt}. It is desirable to maximize the efficiency of the association process. When forming heteronuclear molecules from loosely trapped gases, phase-space density~\cite{hodby05} and the spatial overlap of the two components of a Bose-Fermi mixture have been limiting factors. In an optical lattice, molecules may be formed with almost unit efficiency from sites containing two atoms, and held for long times without the possibility of three-body losses~\cite{thalhammer06}. However, the limit on the number of molecules that may be made with a lattice is the efficiency with which a gas may be loaded into it with exactly two atoms per site.

In this Letter, we show that a two-component gas can be loaded into an optical lattice within a harmonic trap in such a way that each occupied lattice site contains one atom of each species. As well as maximizing molecule production efficiency, our technique will enable the preparation of a lattice with comparatively few defects, if the experimental entropy per particle can be brought to be on the order of the Boltzmann constant, $k_B$. This would allow the formation of dense quantum degenerate gases of dipolar molecules, and the study of interesting new quantum effects where one has both degeneracy and long range, directionally dependent interactions. Such systems could display supersolid phases as well as topological insulating behavior. The system we study here is that of fermionic $^{40}$K and bosonic $^{87}$Rb, which was recently used to produce ultracold dipolar molecules in their absolute ground state~\cite{jila_expt}. Our method, however, is applicable to any two-component gas with a comparatively heavy bosonic species and light fermionic species. We go beyond previous studies~\cite{damski03} by including a realistic trapping potential and the effects of finite temperature.

The experiments of Ref.~\cite{jila_expt} took place in a loose harmonic trap at temperatures close to the Bose-Einstein condensation temperature of $^{87}$Rb, $T_c$.  The magnetic field sweep to produce Feshbach molecules can have near unit efficiency for sufficiently high phase space density~\cite{thalhammer06}, and stimulated Raman adiabatic passage~\cite{jaksch02} (STIRAP) from the Feshbach molecular state to the absolute ground state has been measured to have an efficiency of 92\%~\cite{junye_private}, limited by laser phase coherence. However, the current efficiency for molecule formation in this system ($\sim$20\%) is limited by the initial phase space density that can be attained and by the spatial overlap of the K and Rb clouds~\cite{zirbel08pra}.  
The latter challenge arises because K and Rb have different masses but nearly identical polarizabilities, so the combined optical and gravitational potentials yield a lower trap center for Rb. In addition, while the Rb cloud shrinks as it is cooled below $T_c$, the K cloud remains large due to the Fermi degeneracy pressure.  This further reduces the overlap of the clouds, and also limits the ability to sympathetically cool the fermions.

\begin{table}[t]
\caption{Parameters of the Fermi-Bose Falicov-Kimball model for a K-Rb mixture and different lattice depths. Note that the Rb hopping is more than an order of magnitude less than the K hopping and is ignored. The lattice site $i$ lies at $(ai_x,ai_y)$, with $a$ the lattice constant, and $i_x$ and $i_y$ integers. The trapping potential $V_{i}^{K, Rb}$ corresponds to a trapping frequency of approximately 55\,Hz for the $15E_R^{Rb}$ lattice, and 40\,Hz for the $20E_R^{Rb}$ lattice.}
\label{table: params}
\begin{ruledtabular}
\begin{tabular}{lll}
Parameter&$15E_R^{Rb}$ lattice&$20E_R^{Rb}$ lattice\\
\colrule\T 
$t^K$&$h\,\times$ 190.6~Hz&$h\,\times$ 108.9~Hz\\
$U^{Rb-Rb}$&$h\,\times$ 1086~Hz&$h\,\times$ 1254~Hz\\
$U^{Rb-Rb}/t^K$&5.7&11.515\\$V_i^K=V_i^{Rb}$&$t^K(i_x^2+i_y^2)/121$&$t^K(i_x^2+i_y^2)/121$\\
\colrule\T 
$t^{Rb}$&$h\,\times $ 14.1~Hz&$h\,\times $ 5.4~Hz\\
\end{tabular}
\end{ruledtabular}
\end{table}

One possible approach to achieving a high joint density of K and Rb would be to simply load the gases into a deep optical lattice, so that the Rb atoms form a Mott insulator and the K atoms form a spin-polarized band insulator.  This approach would ensure uniform density of both species at the center of the trap. However, a substantial fraction of both species would remain in the outer region of the trap, where K and Rb have ``metallic'' and superfluid character, respectively, which produce a nonuniform distribution. Furthermore, the sizes of the K and Rb insulating regions will likely not match. Consequently, this approach would not be able to pair all K and Rb atoms, and would leave a severe limitation on the possible efficiency of molecule formation.
Instead, we propose to load the atoms into an optical lattice with the Rb atoms deep in the Mott phase but the K atoms able to hop between sites, and tune the interspecies attraction into a range where the energetically stable phases have {\it precisely one} K {\it and one} Rb atom per lattice site.  In this case, at low temperature, the system will minimize its energy by creating these pre-formed molecules, rather than having only K or Rb atoms on a lattice site.  The system will then be perfectly primed for the formation of ground state molecules with a Feshbach sweep and STIRAP.

We consider optical lattices that are deep enough for the tunnelling of Rb atoms to be negligible.
We may then neglect the kinetic energy and describe the spin-polarized K-Rb mixture with the Fermi-Bose Falicov-Kimball model~\cite{falicov_kimball,ates_ziegler,vollhardt_bdmft},
\begin{eqnarray}
\mathcal{H}&=&-t^K\sum_{ij}c_i^{\dagger}c_j^{}+\sum_i (V_i^K-\mu^K)c_i^{\dagger} c_i^{}\nonumber\\
&+&U^{K-Rb}\sum_i c^\dagger_ic^{}_ib^\dagger_ib^{}_i\label{eq: ham}\\
&+&\sum_i (V_i^{Rb}-\mu^{Rb})b_i^{\dagger} b_i^{}+\frac{1}{2}U^{Rb-Rb}\sum_i b_i^{\dagger} b_i^{}(b_i^{\dagger} b_i^{}-1) \, ,\nonumber
\end{eqnarray}
where $c^\dagger_i$ creates a fermionic $^{40}$K atom at lattice site $i$ and $b^\dagger_i$ creates a bosonic $^{87}$Rb atom at site $i$. The parameters include the K hopping $t^K$, trap potential ($V_i^K$) and chemical potential ($\mu^K$), as well as the corresponding terms for the Rb atoms. 
In addition, we have the attractive interspecies interaction strength, $U^{K-Rb}$, and the repulsive Rb-Rb interaction strength, $U^{Rb-Rb}$. 
The Rb scattering length is 100$a_0$, where $a_0=0.0529$\,nm~\cite{chin_review}. The K-Rb scattering length varies with magnetic field strength $B$ as 
$-191 a_0(1 - \Delta B/[B-B_0])$, where $B_0=546.9$\,G and $\Delta B = -3.1$\,G are the location and width of the resonance employed to make the Feshbach molecules, respectively, and $1\, \mathrm{G} = 10^{-4}$\,T~\cite{chin_review}. The desired $U^{K-Rb}$ is set by choosing the appropriate value of $B$ close to $B_0$.  

We assume a 1030\,nm laser for the optical lattice, which is chosen because it provides approximately the same trapping potential for K and Rb~\cite{ospelkaus06}. 
We work on a two-dimensional lattice with $51\times 51$ lattice sites. The particle number is fixed to be 625 K atoms and 625 Rb atoms in the two-dimensional plane, which is close to half filling of the simulated lattice.
We examine two lattice depths which are both deep in the Mott insulating phase for Rb.  The first is $15E_R^{Rb}$ and the second is $20E_R^{Rb}$, where $E_R^{Rb}= h \times 2.164$~kHz is the recoil energy of the Rb atoms. Assuming this lattice is created by introducing a $40E_R^{Rb}$ potential in the third dimension, the parameters of the Hamiltonian are summarized in Table~\ref{table: params} for the two lattice depths.

\begin{figure}[t]
\centerline{\includegraphics [width=3.2in, angle=0, clip=on]  {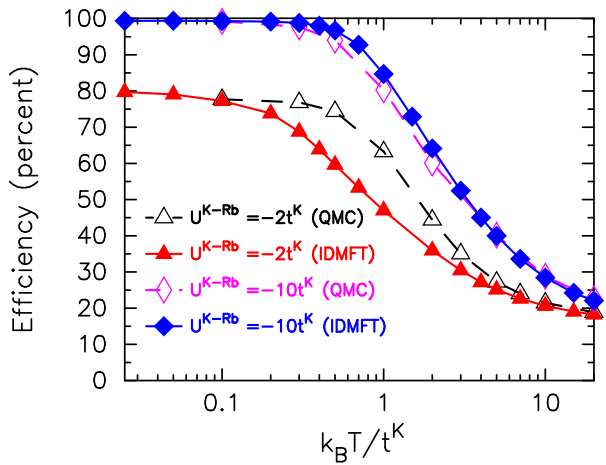}}
\caption[]{
(Color online.) Efficiency for pre-forming molecules on the $15E_R^{Rb}$ lattice for two different $U^{K-Rb}$ interactions. Filled symbols are IDMFT and open symbols are QMC. $U^{K-Rb} = -2t^K$ and $-10t^K$ correspond to $B = 542.9$\,G and 565.9\,G, respectively.
}
\label{fig: eff_vs_t}
\end{figure}

The phenomenon of pre-formed molecules in the Falicov-Kimball model has been widely studied for the case of hard-core bosons (or equivalently spin polarized fermions) on a homogeneous lattice~\cite{fk_molecule_review}. 
In addition, the case of fermions in a harmonic trap was examined in Ref.~\cite{prl_pattern}.
Here we extend this work to examine soft-core bosons in a harmonic trap. We focus on two specific quantities. The efficiency of producing pre-formed molecules measures the joint probability to find  {\it exactly one} K and {\it exactly one} Rb atom on a lattice site. The entropy is employed as an effective temperature scale, under the assumption that the optical lattice is turned on adiabatically and that the lattice does not cause significant heating during the experiment. We have two complementary calculational tools that we employ. Inhomogeneous dynamical mean-field theory (IDMFT)~\cite{idmft} is approximate for a two-dimensional system, but allows us to determine both the production efficiency of pre-formed molecules and the entropy per particle of the mixture.  In contrast, direct quantum Monte Carlo (QMC) simulations~\cite{qmc} are numerically exact, subject to statistics and the possibilities of becoming stuck in local minima, but can only accurately determine the efficiency.  We employed a zero-temperature local-density-approximation analysis to determine the set of parameters that we studied at finite temperatures. 

In Fig.~\ref{fig: eff_vs_t} we plot the efficiency versus temperature for the two different calculational methods on the $15E_R^{Rb}$ lattice. The horizontal axis is on a logarithmic scale to emphasize the differences between the two curves, which disagree only in the narrow region where the efficiency changes rapidly. We note that the disagreement between the two approaches is smaller for deeper lattices. Consequently, the results of Fig.~\ref{fig: eff_vs_t} verify that the IDMFT gives accurate results in the parameter range considered here. Another consistency check we used was to examine the sum rules for the first three moments of the local density of states and for the zeroth and first moment of the imaginary part of the self-energy~\cite{sumrules}. The sum rules are satisfied to high accuracy for all cases except very low temperature, where these functions have sharp peaks that are not easily represented on our coarse frequency grid. This further confirms the quality of the IDMFT approximation.

\begin{figure}[t]
\centerline{\includegraphics [width=3.2in, angle=0, clip=on]  {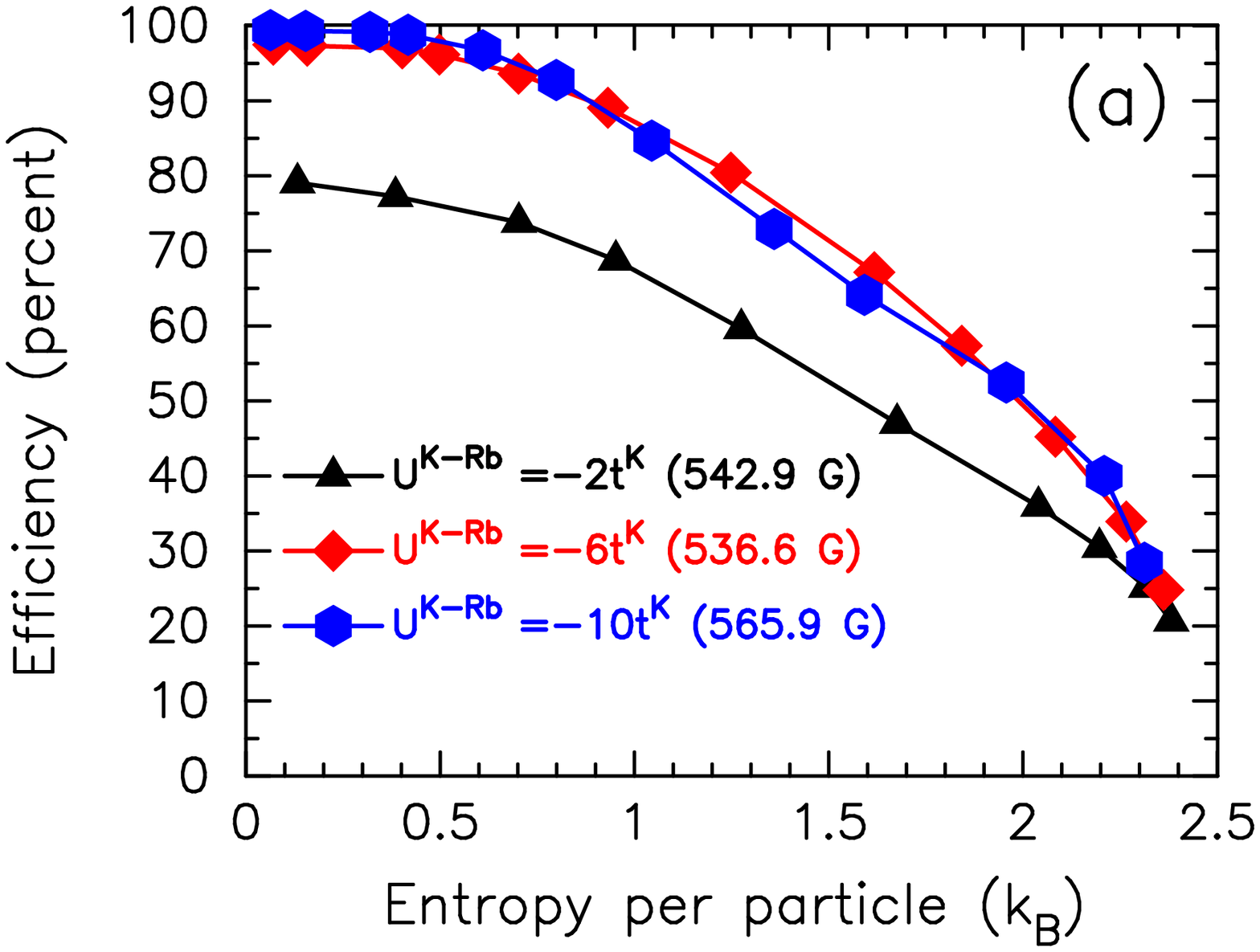}}
\centerline{\includegraphics [width=3.2in, angle=0, clip=on]  {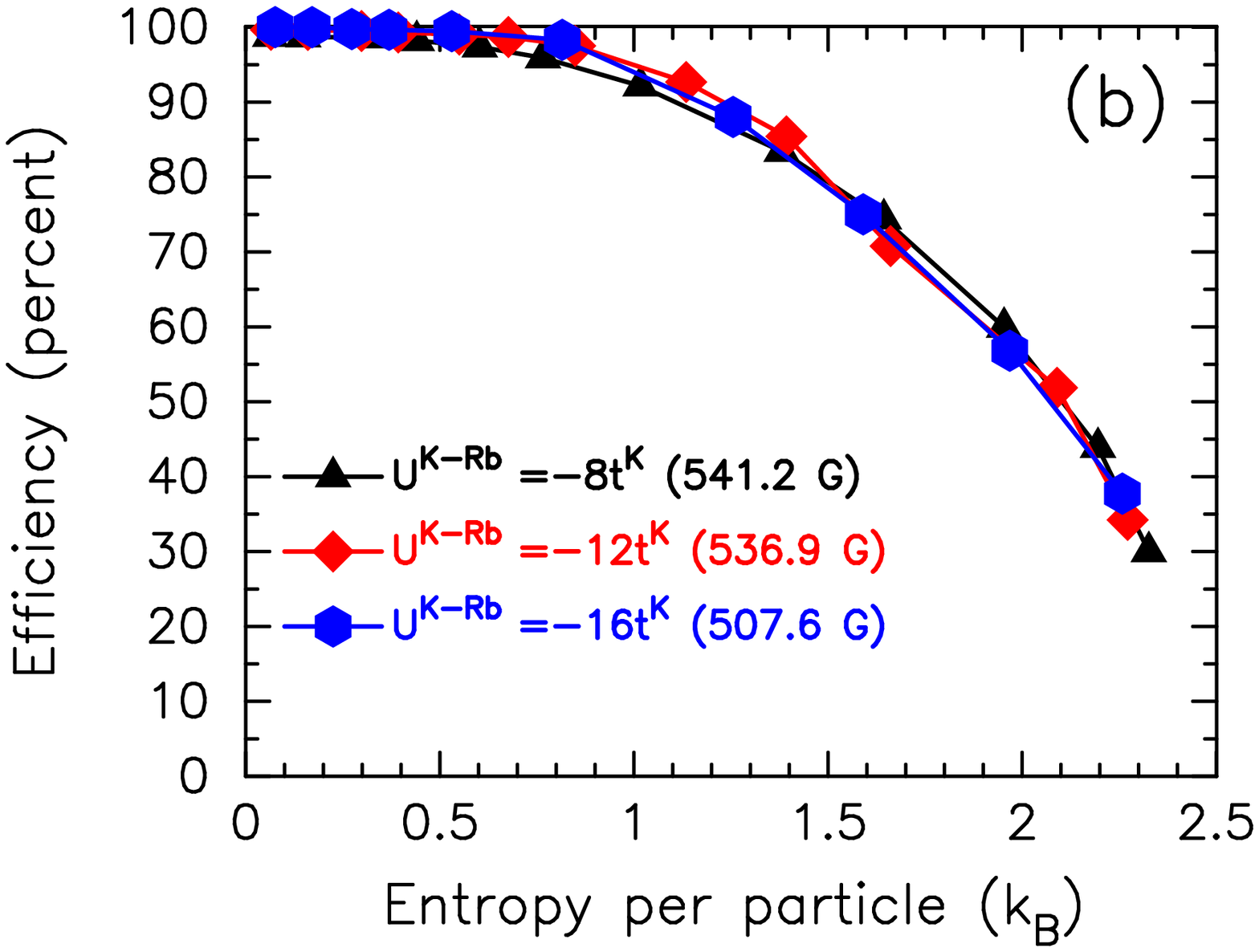}}
\caption[]
{
(Color online.) Efficiency for pre-forming molecules on the (a) $15E_R^{Rb}$ and (b) $20E_R^{Rb}$ lattices versus the entropy per particle, as calculated with IDMFT. The different symbols correspond to different K-Rb interaction strengths.  The efficiency is nearly 100\% at a sufficiently low entropy per particle, corresponding to low temperature, and for large enough K-Rb attraction.
%
}
\label{fig: eff_vs_s}
\end{figure}

In Fig.~\ref{fig: eff_vs_s} we report our main results for efficiency versus entropy per particle at the two different optical lattice depths that we consider. The entropies shown correspond to temperatures ranging from $0.05t^K$ to $10t^K$. In each case we consider three different values of the inter-species interaction strength. All of the values used are weak enough that the number of sites with two Rb atoms, and accompanying three-body recombination loss, are acceptably low.
We note that for a fixed value of the entropy per particle, the deeper lattice will provide a higher efficiency.  For a given lattice depth, once the attraction is strong enough to stabilize the pre-formed molecule, the efficiency depends only weakly on the interaction strength.  By tuning the scattering length into this regime, one can increase the efficiency to 90\% in the shallower lattice, and to more than 95\% in the deeper lattice, even for the relatively high temperature corresponding to an entropy of $k_B$ per particle.

\begin{figure}[h]
\includegraphics[width=3.2in, clip]{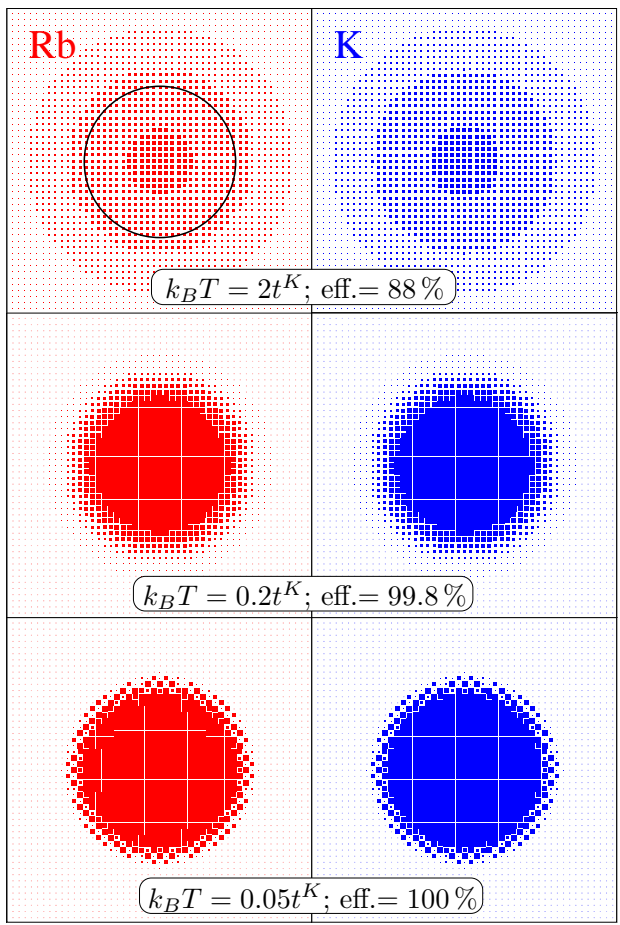}
\caption{(Color online)
Density distributions for Rb (left) and K (right) atoms, at temperatures indicated on the plots along with the resulting efficiency of producing preformed molecules. Here, the lattice has a depth of $20E_R^{Rb}$, and the interspecies attraction is $U^{K-Rb}=-16t^K$.  The symbol size is proportional to the density on each lattice site. Sites inside the solid circle, at the center of the top left panel, show the lattice sites that have a double occupancy of Rb atoms more than 2\% of the time.
}
\label{fig: densities}
\end{figure}

In Fig.~\ref{fig: densities} we plot the densities of the Rb and K atoms for the deeper lattice with $U^{K-Rb}=-16t^K$. The size of the symbols indicate the density on each lattice site.  Note how the Rb and K atoms generally lie on top of each other, but the density approaches one only at the lower temperatures. At higher temperatures, it is more likely that a site could have two Rb atoms. We indicate the temperature at which this begins to become noticeable with the solid circle in Fig.~\ref{fig: densities}, within which sites are doubly occupied more than 2\% of the time. At the lowest temperatures, we can see how the reduced density of the cloud in the outer regions corresponds to nontrivial ordered patterns of the pre-formed molecules, i.e. that the few occupied sites still have one K atom and one Rb atom. One can also see that as the temperature is made too large, there are boundary effects due to the finite size of our system, which restricts the clouds from becoming as large as they should be. This effect overestimates the efficiency and corresponds to the rightmost data in Figs.~\ref{fig: eff_vs_t} and \ref{fig: eff_vs_s}.

In conclusion, the production efficiency of ground state dipolar molecules can be greatly increased by first loading the atomic gases into an optical lattice with an interspecies scattering length such that the system tends to form sites containing one atom of each species. Our scheme works well even for gases prepared at relatively high temperatures, but is predicted to improve as this temperature decreases. Finding the optimal lattice depth, however, requires a number of different choices to be made. Although our calculations predict higher efficiency for deeper lattices, experimental issues such as the longer times required for adiabatic loading require consideration. As the range of scattering lengths within which pre-formed molecules are stable varies with the lattice depth, optimizing these parameters requires further input from experiment.


{\it Acknowledgments:} %
J.~K.~F. and M.~M.~M.~acknowledge support from ARO Grant W911NF0710576 with funds from the DARPA OLE Program. M.~M.~M.~acknowledges Grant No. NN 202 128 736 from Ministry of Science and Higher Education (Poland). P.~S.~J. acknowledges partial support by the U.S. Office of Naval Research. 
Supercomputer time was provided by the ERDC, ASC, and ARSC supercomputer centers of the HPCMP of the DOD via a Challenge grant. Most of the real-axis cold atom calculations were performed during a capabilities application project in 2008 on the Cray XT5 (pingo) at ARSC.

\end{document}